\documentclass[11pt,twoside]{article}
\usepackage{asp2014}

\aspSuppressVolSlug
\resetcounters

\begin{document}

\title{Optimization of the storage database for the Monitoring system of the CTA}

\author{Federico Incardona,$^1$ Alessandro Costa,$^1$ Kevin Munari,$^1$ Pietro Bruno,$^1$ Stefano Germani,$^2$ Alessandro Grillo,$^1$ Igor Oya,$^3$ Dominik Neise,$^3$ and Eva Sciacca$^1$ for the CTA Observatory}
\affil{$^1$INAF, Osservatorio Astrofisico di Catania, Via S Sofia 78, I-95123 Catania, Italy; \email{federico.incardona@inaf.it}}
\affil{$^2$Universit\`a di Perugia, Dipartimento di Fisica e Geologia, Italy}
\affil{$^3$CTA Observatory gGmbH}

\paperauthor{Federico Incardona}{federico.incardona@inaf.it}{}{INAF}{Osservatorio Astrofisico di Catania}{Catania}{}{I-95123}{Italy}
\paperauthor{Alessandro Costa}{alessandro.costa@inaf.it}{}{INAF}{Osservatorio Astrofisico di Catania}{Catania}{}{I-95123}{Italy}
\paperauthor{Kevin Munari}{kevin.munari@inaf.it}{}{INAF}{Osservatorio Astrofisico di Catania}{Catania}{}{I-95123}{Italy}
\paperauthor{Pietro Bruno}{pietro.bruno@inaf.it}{}{INAF}{Osservatorio Astrofisico di Catania}{Catania}{}{I-95123}{Italy}
\paperauthor{Stefano Germani}{stefano.germani@unipg.it}{}{Universit\`a di Perugia}{Dipartimento di Fisica e Geologia}{Perugia}{}{I-06123}{Italy}
\paperauthor{Alessandro Grillo}{alessandro.grillo@inaf.it}{}{INAF}{Osservatorio Astrofisico di Catania}{Catania}{}{I-95123}{Italy}
\paperauthor{Igor Oya}{igor.oya@cta-observatory.org}{}{CTA Observatory gGmbH}{}{}{}{}{}
\paperauthor{Dominik Neise}{neised@phys.ethz.ch}{}{CTA Observatory gGmbH}{}{}{}{}{}
\paperauthor{Eva Sciacca}{eva.sciacca@inaf.it}{}{INAF}{Osservatorio Astrofisico di Catania}{Catania}{}{I-95123}{Italy}



  
\begin{abstract}

We present preliminary test results for the correct sizing of the bare metal hardware that will host the database of the Monitoring system (MON) for the Cherenkov Telescope Array (CTA). The MON is the subsystem of the Array Control and Data Acquisition System (ACADA) that is responsible for monitoring and logging the overall CTA array. It acquires and stores monitoring points and logging information from the array elements, at each of the CTA sites. MON is designed and built in order to deal with big data time series, and exploits some of the currently most advanced technologies in the fields of databases and Internet of Things (IoT). To dimension the bare metal hardware required by the monitoring system (excluding the logging), we performed the test campaign that is discussed in this paper. We discuss here the best set of parameters and the optimized configuration to maximize the database data writing in terms of the number of updated rows per second. We also demonstrate the feasibility of our approach in the frame of the CTA requirements.
  
\end{abstract}

\section{Introduction}
The Cherenkov Telescope Array \citep{ACHARYA20133} will be the largest and most advanced ground-based observatory for detection of electromagnetic radiation between 20 GeV and 300 TeV. It will be composed of several tens of telescopes installed on two arrays, one in the northern hemisphere (Canary Islands, Spain) and one in the southern hemisphere (Paranal, Chile). Together with the scientific data produced by CTA, a big volume of housekeeping and auxiliary data coming from weather stations, instrumental sensors, logging files, etc., must be collected as well. More precisely, we expect to collect information from about 200.000 monitoring points, sampled between 1 and 5 Hz, for a maximum data rate for writing operations of 26 Mbps. We designed the Monitoring system (MON) of CTA in order to deal with big data time series, and to make those data immediately available for the operator interface and for quick-look quality checks, as well as to store them for later detailed inspection \citep{Costa:2021c8}. To handle the storage of the whole amount of data, the MON makes use of Apache Cassandra\footnote{\url{https://cassandra.apache.org}.}, an open source NoSQL distributed database that is able to store extremely high data volume at relatively high rate, and while being fault tolerant. In this paper we present some tests we performed in order to dimension the bare metal hardware required for the Monitoring system (excluding the Logging).

\section{Storage Database characteristics}
We based the storage of MON on Apache Cassandra 3.11.9. The operations of reading and writing are performed by Cassandra using a ``Primary Key'' on a table that is composed, in turn, by a ``Partition Key'', which defines a unique set of rows that is managed within a node of the cluster, and an optional ``Clustering Key'', which handles the data arrangement part. In the example shown in Figure \ref{cassandra_table}, partitions are univocally determined by a unique pair of \textit{Sensor} and \textit{Date}, while the associated list of \textit{Timestamp}s defines the number of rows in such partition.
\articlefigure[scale=0.21]{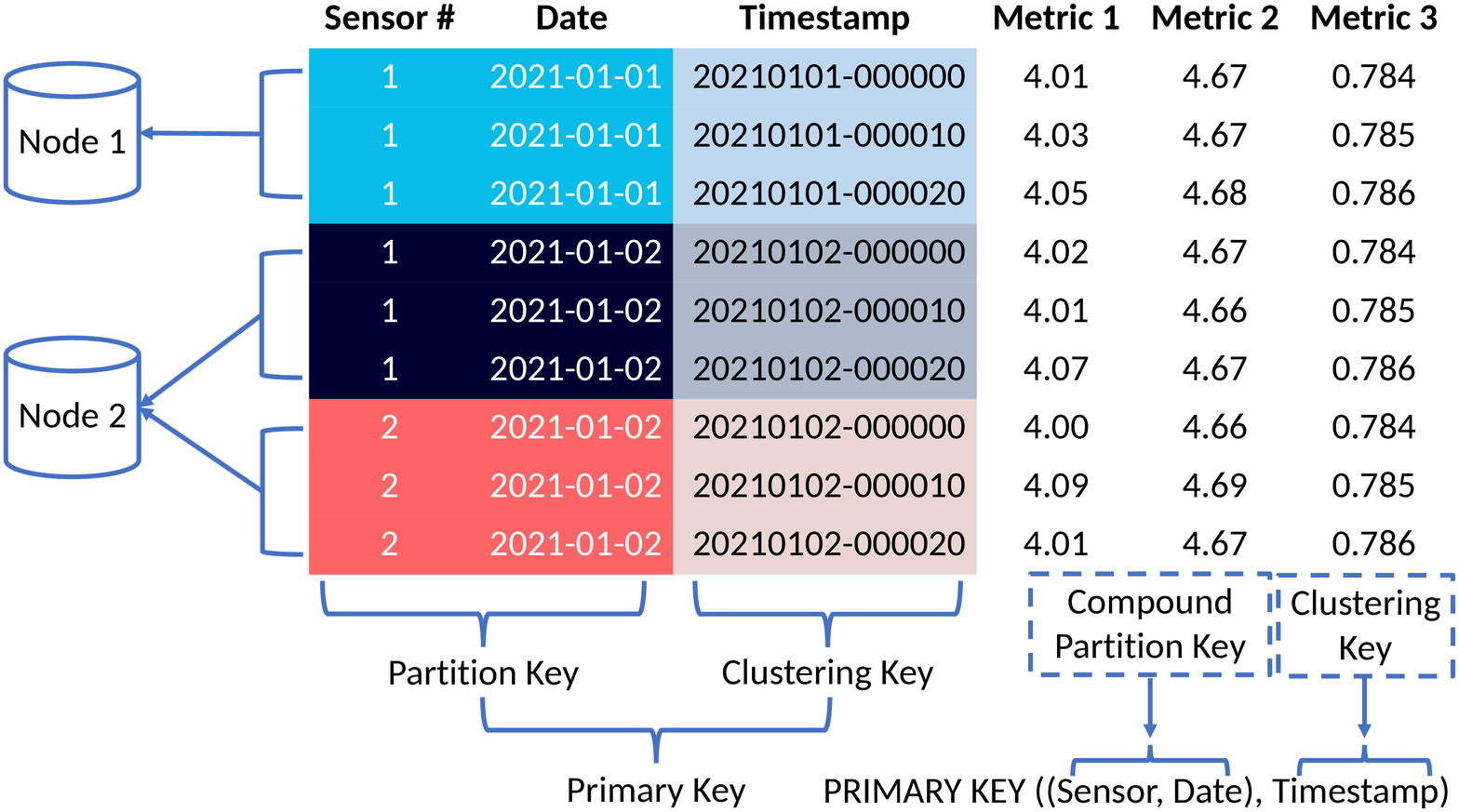}{cassandra_table}{Cassandra partition system. (From \url{https://www.instaclustr.com}).}

Cassandra provides the so-called ``Cassandra-stress'', which is a Java-based stress testing utility that can be used to perform basic benchmarking and load testing. For each field of the data model, Cassandra-stress allow the user to simulate the data entries, according to five distributions: fixed, Gaussian, uniform, exponential and ``extreme''. Furthermore, it is also possible to specify the distributions of the \textit{size} and the \textit{population} of that field. For the Clustering Key, a further distribution is allowed, the \textit{cluster} distribution, which specifies the distribution of the number of rows in the partitions.

\section{Simulations}
In the wake of following \citet{6253535}, we wanted to provide an estimation of the writing performance of our data model with Cassandra. To this purpose, we performed several simulations in the bare metal by exploiting Cassandra-stress. We used a single machine equipped with two Intel(R) Xeon(R) CPU E5-2650 v3 at 2.30GHz and 40 CPUS (20 physical, 40 in multithreading), 256 GB of RAM, SATA hard disks with 64 MB cache and 7200 rpm (6Gb/s) in RAID5, and RAID controller Series 6 - 6G SAS/PCle 2 clock at 33 MHz.

\subsection{Data model assumptions}
In our data model a monitoring value is defined by eight fields. The \textit{assembly} is the name of the component where the property is exposed. It should correspond to the monitoring data source, e.g. ``PowerSupply1'', and it is characterized by a \textit{serial\_number}. To each property to be monitored it is associated a \textit{name}, e.g. ``Current'', and a \textit{unit}, e.g. ``A''. The time at which such property is produced by the device is called \textit{source\_timestamp}, while the one at which it is received by the server is called \textit{server\_timestamp}. In the context of our simulations, these times could assume every value in the whole range of the ``long'' type. The \textit{env\_id} is a string representing the environment, e.g. ``CTA-N''. Finally, the time-series \textit{data} are collected in the format of a list of double, which represents the status or quality of the property. For each field of our data model we assumed the distributions described in Table \ref{table_distrib}. As for the \textit{data} field, there is no population distribution since it may refer to different kinds of observables, while the size distribution is exponential since most of them are small, decreasing exponentially.

\begin{table}[!ht]
\caption{Distributions used in the simulations.}\label{table_distrib}
\smallskip
\begin{center}
{\small
\begin{tabular}{llll}  
\tableline
\noalign{\smallskip}
Field & Population (\#) & Size (bytes) & Cluster: (\#)\\
\noalign{\smallskip}
\tableline
\noalign{\smallskip}
\textit{assembly} & Uniform: 1 $\div$ 4.000 & Gaussian: 12 $\div$ 24  & \\
\noalign{\smallskip}
\textit{name} & Uniform: 1 $\div$ 100.000 & Gaussian: 8 $\div$ 16  & \\
\noalign{\smallskip}
\textit{serial\_number} & Uniform: 1 $\div$ 4.000 & Fixed: 12  & \\
\noalign{\smallskip}
\textit{server\_timestamp} & Gaussian: 1 $\div$ 1.000.000.000 & Fixed: 8  &  Fixed: $r$ \\
\noalign{\smallskip}
\textit{source\_timestamp} & Gaussian: 1 $\div$ 1.000.000.000 & Fixed: 8  & \\
\noalign{\smallskip}
\textit{units} & Uniform: 1 $\div$ 20 & Gaussian: 1 $\div$ 4  & \\
\noalign{\smallskip}
\textit{env\_id} & Uniform: 1 $\div$ 150 & Gaussian: 3 $\div$ 10  & \\
\noalign{\smallskip}
\textit{data} &  & Exponential: 1 $\div$ 100  & \\
\noalign{\smallskip}
\tableline\
\end{tabular}
}
\end{center}
\end{table}

The table associated to our data model is represented in Figure \ref{data_model}. We use a compound Partition Key made by \textit{assembly} and \textit{name}, and a Clustering Key made by the \textit{server\_timestamp} alone. 
\articlefigure[scale=0.2]{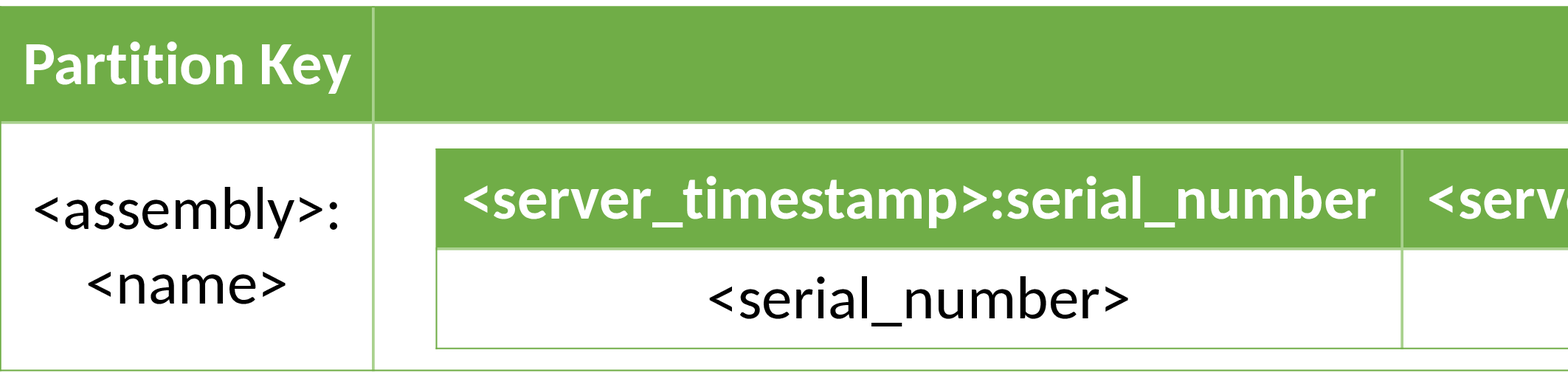}{data_model}{Cassandra table associated to a monitoring value.}

\section{Results}
We ran some tests to provide an estimation of the writing performance of our data model with Cassandra, in terms of the rows writing rate. In each test, we repeated the simulation of 1.000.000 insert batches by increasing, each time, the number of parallel threads exploited, in the range allowed by Cassandra-stress for that test. We performed several tests for different values of $p$, the number of partitions per insert batch, and the number of rows per partition $r$ (corresponding to the value of the fixed cluster distribution for the \textit{server\_timestamp}, see Table \ref{table_distrib}).

As shown in Fig. \ref{resultsplot}, the cases of $p$ = 1, $r$ = 100 and $p$ = 1, $r$ = 1000 maximize the writing performance, which reaches the level of 700.000 updated rows per second, corresponding to a data throughput of about 18 Mbps. This value is narrowly below the expected data throughput of 26 Mbps for the monitoring of CTA. We notice also that the data writing benefits of multithreading from 40 threads on, and that we obtain low performance when the product of $p$ and $r$ is greater than 10.000. 
\articlefigure[trim={0 0 0 1.5cm}, clip, scale = 0.4]{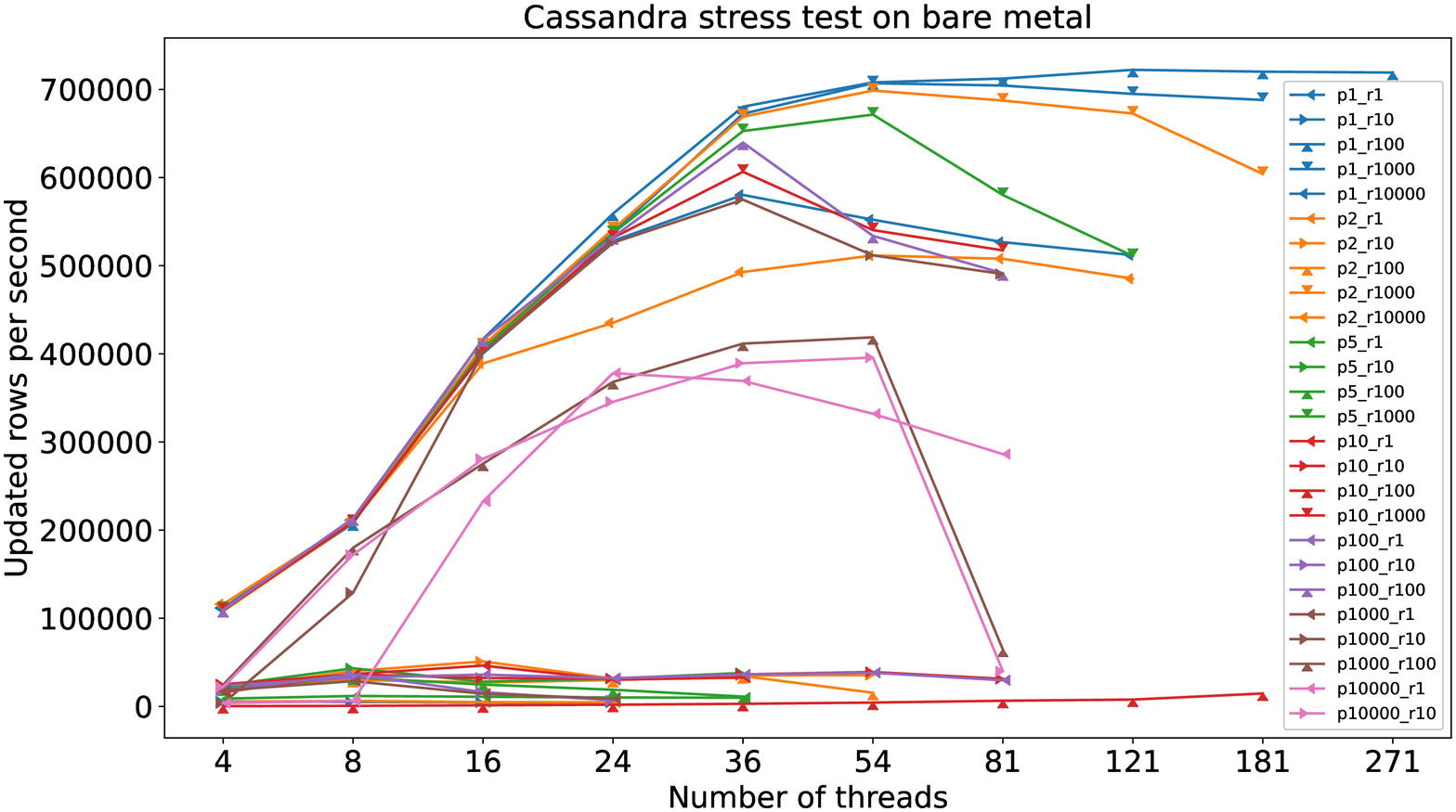}{resultsplot}{Results of the performance tests of our data model with Cassandra. Each line represents the writing rate for a given simulation as a function of the number of parallel threads exploited. Simulation parameters are the number of partitions per insert batch ($p$, different colors) and of rows per partition ($r$, different markers).}

\section{Conclusions}
Our results show that CTA will require two machines with the capabilities of our test machines to address the storage of its Monitoring system. We aim to perform further tests with different workloads (reads and mixed), and a not-fixed number of updated partitions and rows per batch. Furthermore, we plan to measure the performance of our data model with Cassandra in a virtual environment and by exploiting a cluster.


\bibliography{X2-002}


\end{document}